%% file: apssamp.tex
\documentclass[%
 reprint,
 amsmath,amssymb,
 aps,
]{revtex4-1}

\usepackage{lineno}  % for line numbering during review
\usepackage{graphicx}% Include figure files
\usepackage{dcolumn}% Align table columns on decimal point
\usepackage{bm}% bold math
\begin{document}

%\preprint{APS/123-QED}

\title{Rare decays at LHCb}%

\author{Francesco Polci}
% \altaffiliation[Also at ]{Physics Department, XYZ University.}
\affiliation{LPNHE, Universit\'{e} Pierre et Marie Curie, Universit\'{e} Paris Diderot, CNRS/IN2P3, Paris, France}
\collaboration{on behalf of the LHCb Collaboration}

%\affiliation{Second institution and/or address}
%\affiliation{Third institution}

\date{\today}% It is always \today, today,

\begin{abstract}
The results on rare decay processes obtained by the LHCb experiment  using 1.0 $fb^{-1}$
of $pp$ collisions collected  in 2011 at a centre-of-mass energy of $\sqrt{s}$=7 $TeV$ are presented. 
Branching fractions, angular distributions, CP and isospin asymmetries are investigated to search for new physics effects.
\end{abstract}

\maketitle

%\tableofcontents
% %%%%%%% CHOOSE --------
%% ----------------------------------
%% Line numbering on the left margin
%% ----------------------------------
%% Uncomment during review phase.
%% Comment it out before a final submission.
%\linenumbers

\section{\label{sec:introduction} Introduction}
The LHCb experiment  performs indirect searches of new physics effects, in a complementary way with respect to the direct searches performed by other experiments. In this respect, rare decays offer a perfect framework. They are suppressed or forbidden in the standard model (SM), so the presence of new physic (NP)  could sensibly modify their properties, including branching fractions, angular distributions of the decay products, CP and isospin asymmetries. They mostly occur via loop diagrams where new particles could show up, with masses up to higher scales than those directly accessible. In case of agreement with the SM predictions,  the results  set constraints on the Wilson coefficients of the effective Hamiltonian  describing the possible new physics contributions to the process under study.

The LHCb experiment~\cite{Alves:2008zz} is well adapted to perform these searches. It takes best profit of the large $b$$\bar{b}$ and $c$$\bar{c}$ cross-sections at the  LHC thanks to the high trigger efficiency, the highly performing particle identification system, as well as the excellent tracking system, which allows for a precise determination of secondary vertices, impact parameters and momenta.

The analyses presented here rely on 1.0 $fb^{-1}$ of $pp$ collisions collected by the LHCb experiment in 2011 at a centre-of-mass  energy of $\sqrt{s}=7$ $TeV$. This data sample  allow to push further upper limits on the branching fractions, get the first evidence of very rare decays and to perform a detailed study of the properties of those decays for which a sizeable sample is collected.

The analyses approach, though with differences specific to each decay for which we refer to the publications, follows some general lines.
 The background from random combination of tracks in the event, called combinatorial background, is suppressed using multivariate techniques exploiting the discriminating power of the kinematic variables. Specific backgrounds peaking in the signal region are reduced with dedicated vetoes. Whenever possible, control channels are used in order to rely as  less as possible on simulation. The measurement of the branching fractions is performed by  normalizing to decays of well known branching fractions, in order to get rid of the uncertainties in the knowledge of the luminosity and the  $b$$\bar{b}$ and $c$$\bar{c}$ cross-sections. The uncertainties on the measured properties of the decays are often statistically dominated. Finally, when the decay is not observed, the CLs method~\cite{Read:2002hq} is used as statistical approach to set upper limits on the branching fractions.

\section{\label{sec:level2}Very rare decays searches}
%Ks->mumu
The $K_s\to \mu^+ \mu^-$ decay is strongly suppressed in the SM, with an expected branching fraction $BR(K_s\to \mu^+ \mu^-)=(5.0 \pm 1.5)\times 10^{-12}$~\cite{Isidori:2003ts}. LHCb has about $10^{13}$ $K_s$ decays per $fb^{-1}$ in its acceptance, and this sample allows to improve the current limit on the branching fraction of this decay. The $K_s \to \pi^+ \pi^-$ channel, which has a similar kinematics, is used to train the boosted decision tree (BDT) used for the selection. This decay is also used  as a normalization channel. The backgrounds are of combinatorial type and from $K_s \to \pi^+ \pi^-$ decays when the pions are misidentified as muons. Candidates are classified in bins of BDT. In each bin the  number of events is compared to signal and background expectations. Though no signal is observed, LHCb is able to set an upper limit of $BR(K_s\to \mu^+ \mu^-)< 9 \times 10^{-9}$ at 90$\%$ $C.L.$ ($< 11 \times 10^{-9}$ at 95$\%$ $C.L.$),   thirty times better than the results of previous experiments~\cite{Aaij:2012rt}.

%tau->3mu
The $\tau \to \mu^+  \mu^+ \mu^-$ is a lepton flavor violating process. The $\tau$ production at LHCb is of about 80 $\mu b$, and it comes mainly from $D_s$ ($\sim80\%$) and $B$ ($\sim20\%$) decays. Having three muon in the final state, a very clear signature is expected, with background coming essentially from combinatorial and $D_s \to \eta(\mu\mu\gamma)\mu \bar{\nu_{\mu}}$ decays. The $D_s\to \phi(\mu\mu)\pi$ channel is used as control and normalization sample.  The preliminary limit is $BR(\tau \to \mu^+  \mu^+ \mu^-)<7.8 \times 10^{-8}$ at $95\%$ confidence level $(C.L.)$ ($<6.3 \times 10^{-8}$ at $90\%$ C.L.)~\cite{tauto3mu}.

%B->4mu
The $B^0_{(s)} \to \mu^+\mu^-\mu^+\mu^-$ decays are strongly suppressed in the SM. They contain two contribution: a resonant one, of the order of (2.3$\pm$0.9)$\times10^{-8}$ in the SM~\cite{PDG},  where two muons of opposite charges come from a $J/\psi$ and the other two from a $\phi$ decay; and a non resonant one expected to be lower than $10^{-10}$~\cite{Melikhov:2004mk}. In the MSSM models those branching fractions are enhanced by the presence of a scalar $S$ and pseudo-scalar $P$ sgoldinos, each decaying into muon pairs. In the LHCb analysis the resonant component is removed and used as a control sample for the estimation of the non resonant one. The branching fractions are normalized to the $B^0 \to J/\psi K^*$ decay. The  background is fully dominated by the combinatorial component. The resulting upper limits, in the SM framework, are: 
$BR(B^0_s\to  \mu^+\mu^-\mu^+\mu^-) < 1.2 \times 10^{-8}$ at $90\%$ C.L. ($< 1.6 \times 10^{-8}$ at $95\%$ C.L.) and 
$BR(B^0    \to  \mu^+\mu^-\mu^+\mu^- )< 5.3 \times 10^{-9}$ at $90\%$ C.L. ($< 6.6 \times 10^{-9}$ at $95\%$ C.L.). 
For the MSSM, assuming the masses of the new sgoldinos to be $m_S=2.5$ $GeV/c^2$ and $m_P=214$ $MeV/c^2$ from the Hyper CP collaboration, we measure:
$BR(B^0_s\to S(\mu^+\mu^-)P(\mu^+\mu^-)) < 1.2 \times 10^{-8}$ at $90\%$ C.L. ($< 1.6  \times 10^{-8}$ at $95\%$ C.L.) and 
$BR(B^0    \to S(\mu^+\mu^-)P(\mu^+\mu^-)) < 5.1 \times 10^{-9}$at $90\%$ C.L.  ($< 6.3  \times 10^{-9}$ at $95\%$ C.L.)~\cite{BTo4mu} . 

%Bs->mumu
%The $B_s\to \mu\mu$ decay is discussed elsewhere but start to give a sizable signal. 

%D decays
$D$ decays are also interesting probes of  new physics. The $D_{(s)}^+\to \pi^+ \mu^+ \mu^-$ decays gives highlights on the flavor changing neutral current (FCNC) $c\to u$ processes. A similar analysis searching for  $D_{(s)}^+\to \pi^- \mu^+ \mu^+$, a decay forbidden in the SM, test the lepton-number violation as well as the possibility of neutrino being Majorana particles. These analyses use the $D^+_{(s)}\to \pi^+ \phi(\mu^+ \mu^-)$ as control channel. Since no signal is observed, upper limits are set, which turn out to be from   about fifthy to hundred times better than previous measurements~\cite{DToPimumu} .

\section{\label{sec:level2}Analysis of the $b\to s(d)$ FCNC properties}
The FCNC $b \to s(d)ll$ transitions are forbidden at tree level in the SM and proceed only via electroweak penguin and box diagrams. Possible new physics contributions could arise in the loops from right handed current and new scalar or pseudo-scalar operators. It is a rich category of decays. For many of them we have collected enough data to study many observables.  

\begin{figure}
\includegraphics[width=8.4cm]{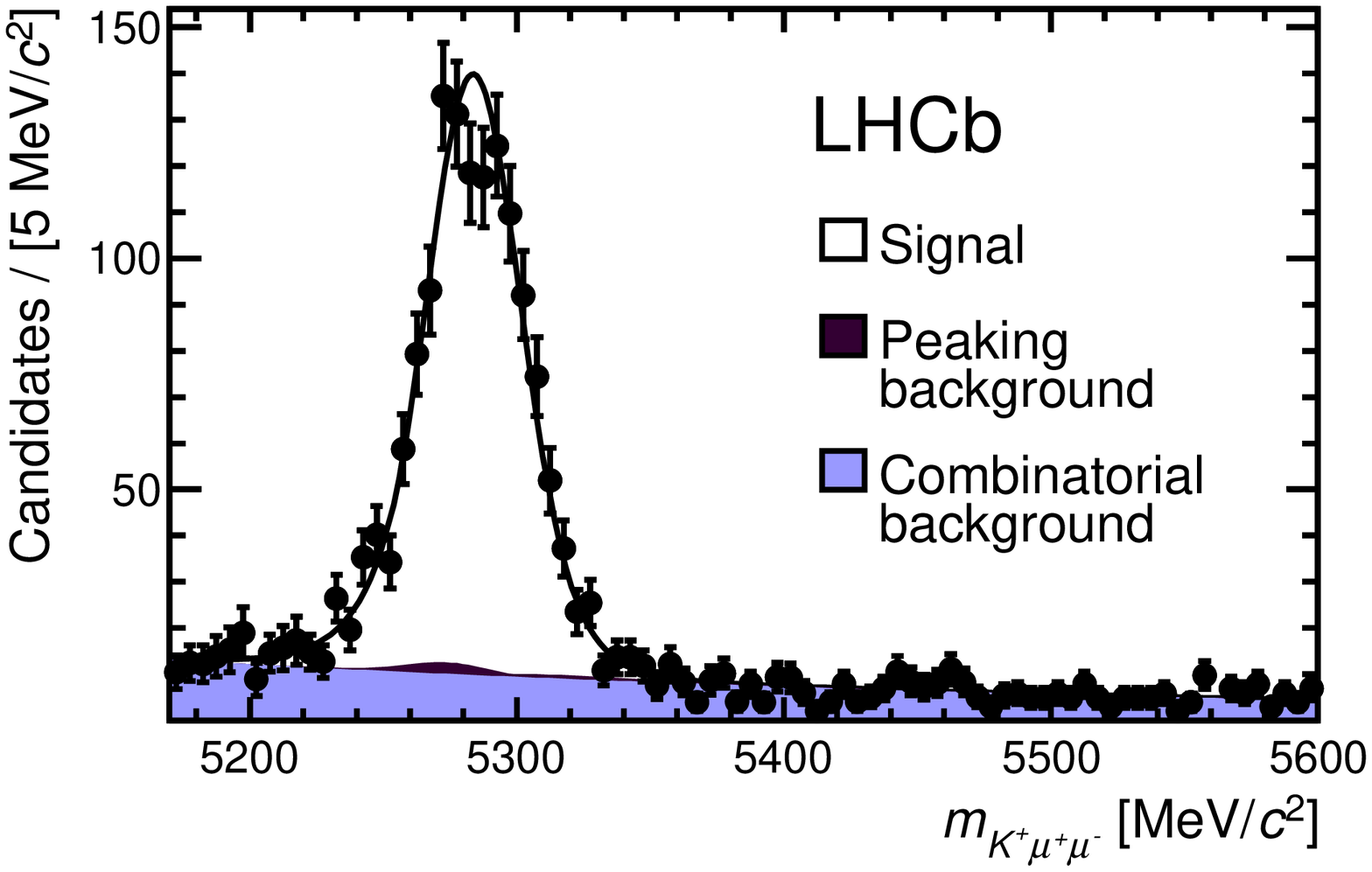} \\
%\end{figure}
%\begin{figure}
\includegraphics[width=8.4cm]{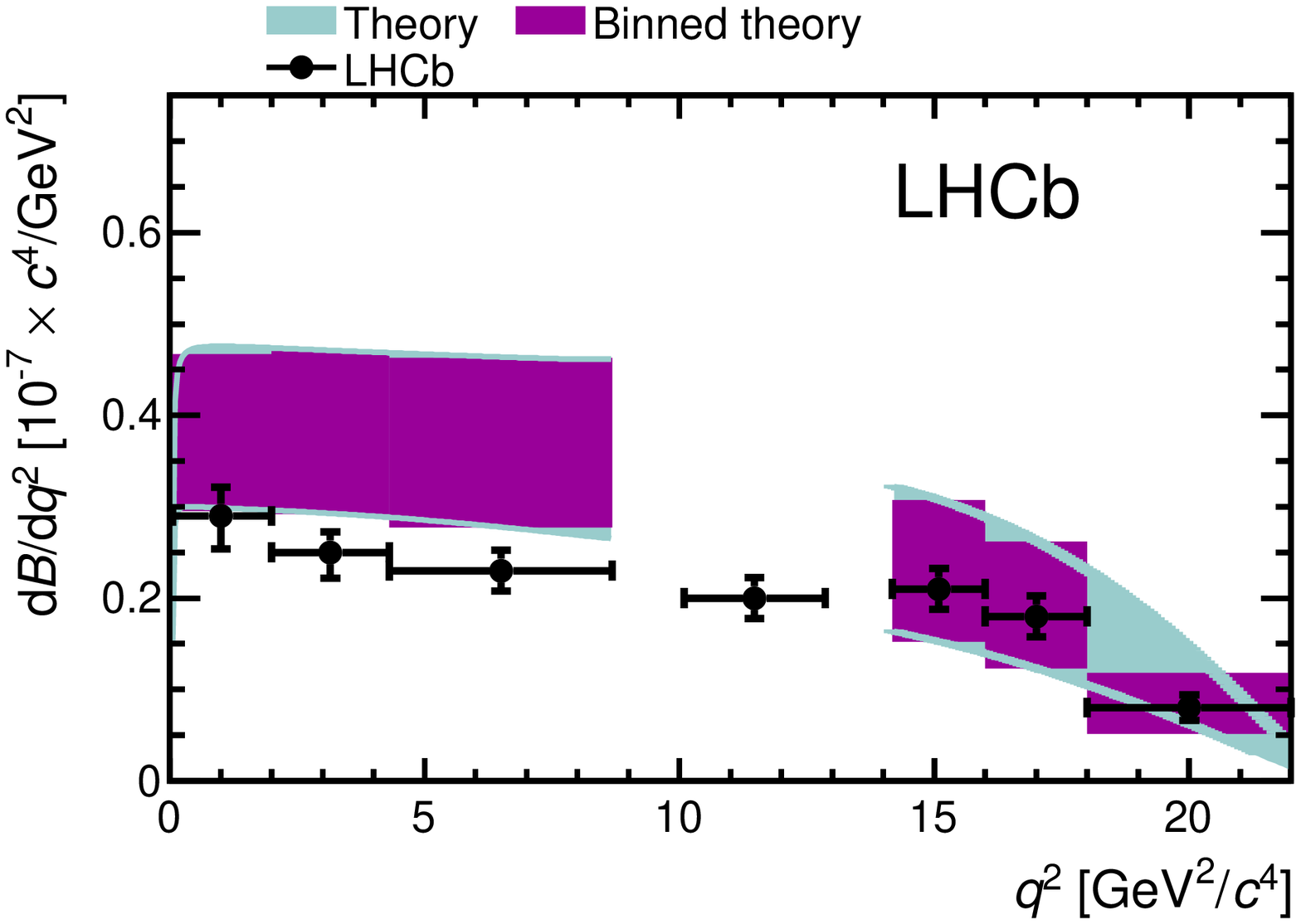}\\ % Here is how to import EPS art
%\end{figure}
\includegraphics[width=4.2cm]{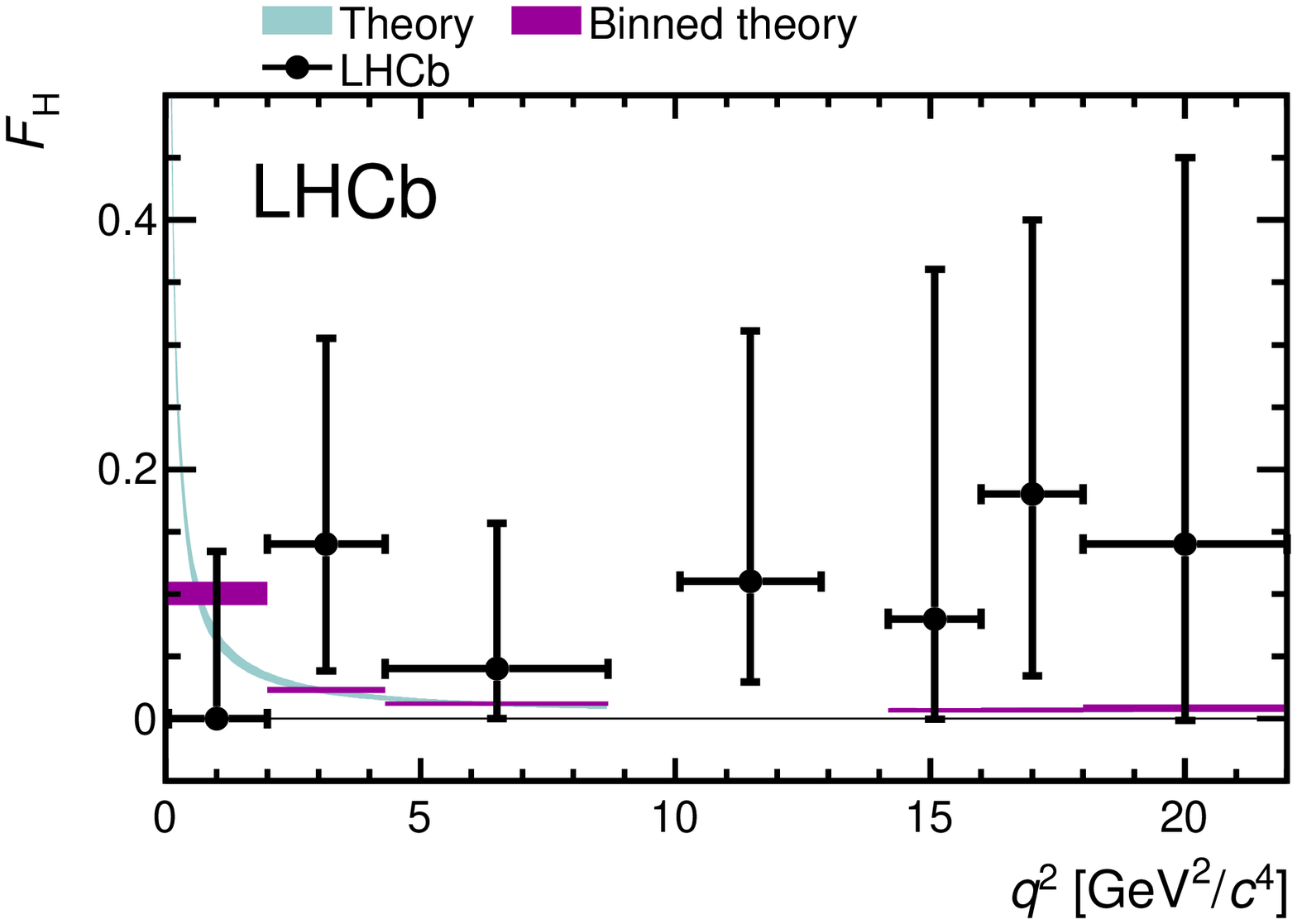}
\includegraphics[width=4.2cm]{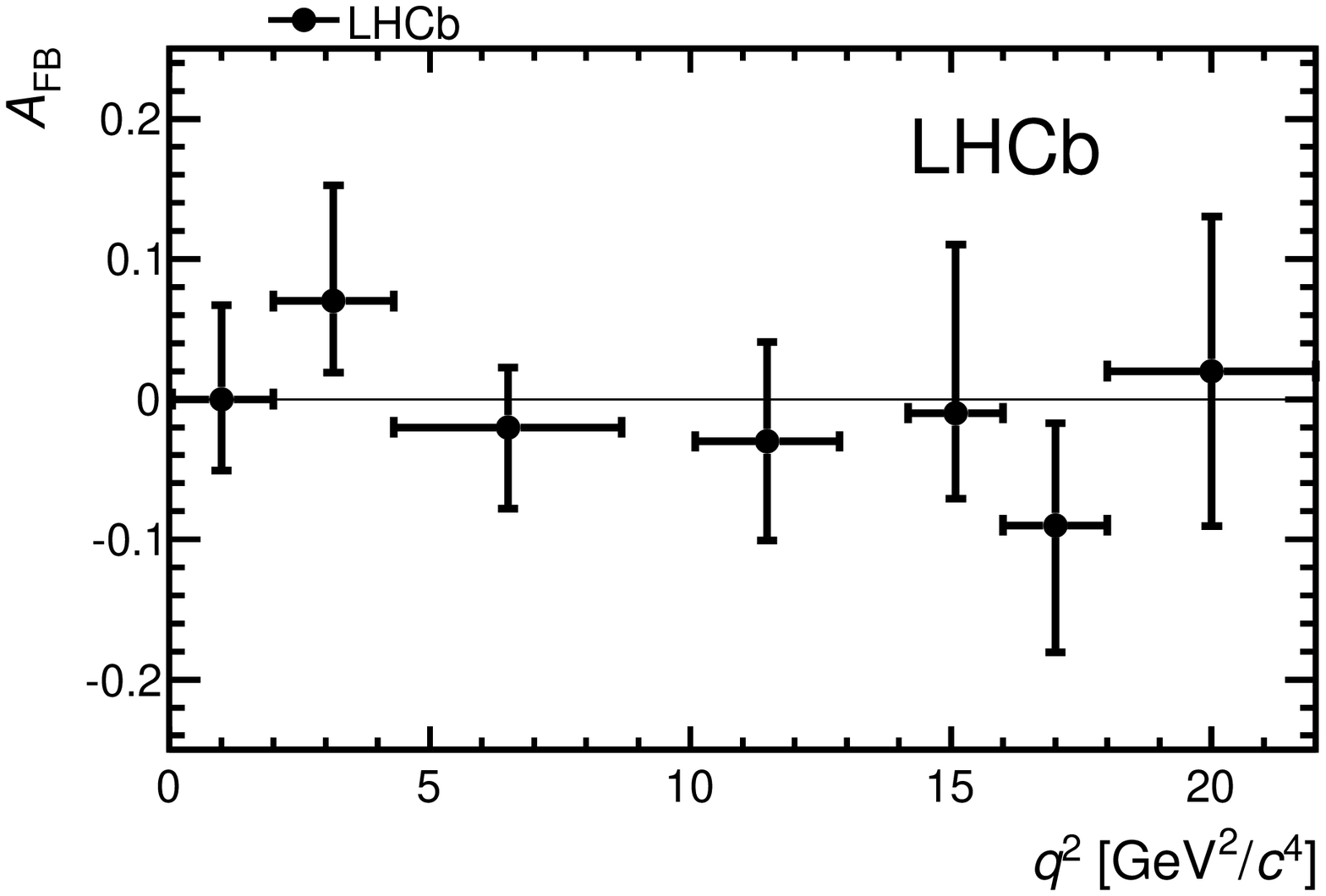}
\caption{\label{fig:BToKmumu} Invariant mass $m_{K^+ \mu^+\mu^-}$ of the $B^+ \to K^+ \mu^+\mu^-$ candidates and results of the measurements of the $BR$, $F_H$ and $A_{FB}$ as function of $q^2$ (black dots) compared to the theoretical predictions (colored curves).}
\end{figure}

%B->Kmm
The $B^+ \to K^+ \mu^+\mu^-$ decay is a $b \to s$ transition that is analyzed by LHCb in the range $0.05 < q^2 < 22$ $GeV^2/c^4$, where $q^2$ is the di-lepton invariant mass squared.  
The decay $B^+ \to K^+ J/\psi(\mu^+\mu^-)$  has been used both to train the BDT used in the selection and as normalization channel.  A clear signal is observed with a small residual background essentially due to combinatorial and $B^+ \to K^+ \pi^+\pi^-$ and $B^+ \to \pi^+ \mu^+\mu^-$ decays.
The differential branching fraction as function of $q^2$ is measured, as well as the total branching fraction: $BR(B^+ \to K^+ \mu^+\mu^-) = (4.36 \pm 0.15 \pm 0.18) \times 10^{-7}$. An angular analysis of the decay products of the distribution as function of the helicity angle of the muon $\theta_L$  is performed and allows to measure the forward-backward asymmetry $A_{FB}$, expected to be zero in the SM, as well as the $F_H$ parameter. The measurements are, at the moment, compatible with the SM expectations. The results~\cite{Aaij:2012vr} are shown in figure~\ref{fig:BToKmumu}.    

\begin{figure}
\includegraphics[width=8.4cm]{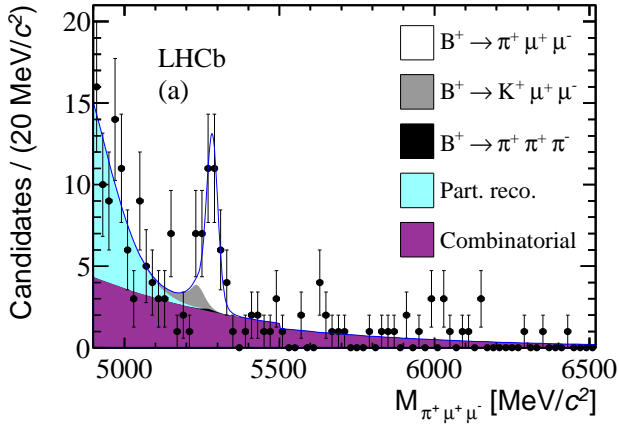} 
\caption{\label{fig:BToPimumu}  Invariant mass $m_{\pi^+ \mu^+\mu^-}$ of the $B^+ \to \pi^+ \mu^+\mu^-$ candidates }
\end{figure}

%B->Pimumu
The $B^+ \to \pi^+ \mu^+\mu^-$ decay is a $b \to d$ transition, predicted to be $(2.0 \pm 0.2) \times 10^{-8}$ in the SM~\cite{Wang:2007sp}. Combined with the $B^+ \to K^+ \mu^+\mu^-$ decay, it  provides a measurement of $|V_{td}|/|V_{ts}|$ alternative to the current ones coming from the radiative decays and the mixing processes. Here again  $B^+ \to K^+ J/\psi(\mu^+\mu^-)$ is used as control sample and normalization channel. The simultaneous fit to the invariant mass distribution of the  $B^+ \to \pi^+ \mu^+\mu^-$ and $B^+ \to K^+ \mu^+\mu^-$ candidates  allows to observe for the first time a $b\to dll$ transition with 5.2 $\sigma$ significance, as shown on figure~\ref{fig:BToPimumu}. The branching fraction is determined to be $BR(B^+ \to \pi^+ \mu^+\mu^-)=(2.3 \pm 0.6 \pm 0.1) \times 10^{-8}$.  
The measured ratio of the two decays is  
$R= BR(B^+ \to \pi^+ \mu^+\mu^-)$/$BR(B^+ \to K^+ \mu^+ \mu^- ) = 0.053 \pm 0.014 \pm 0.001$,  
giving $|V_{td}|/|V_{ts}| = 0.266\pm0.035\pm0.003$ in agreement with the radiative decays and mixing processes measurements~\cite{LHCb:2012de}.  

\begin{figure}
\includegraphics[width=8.4cm]{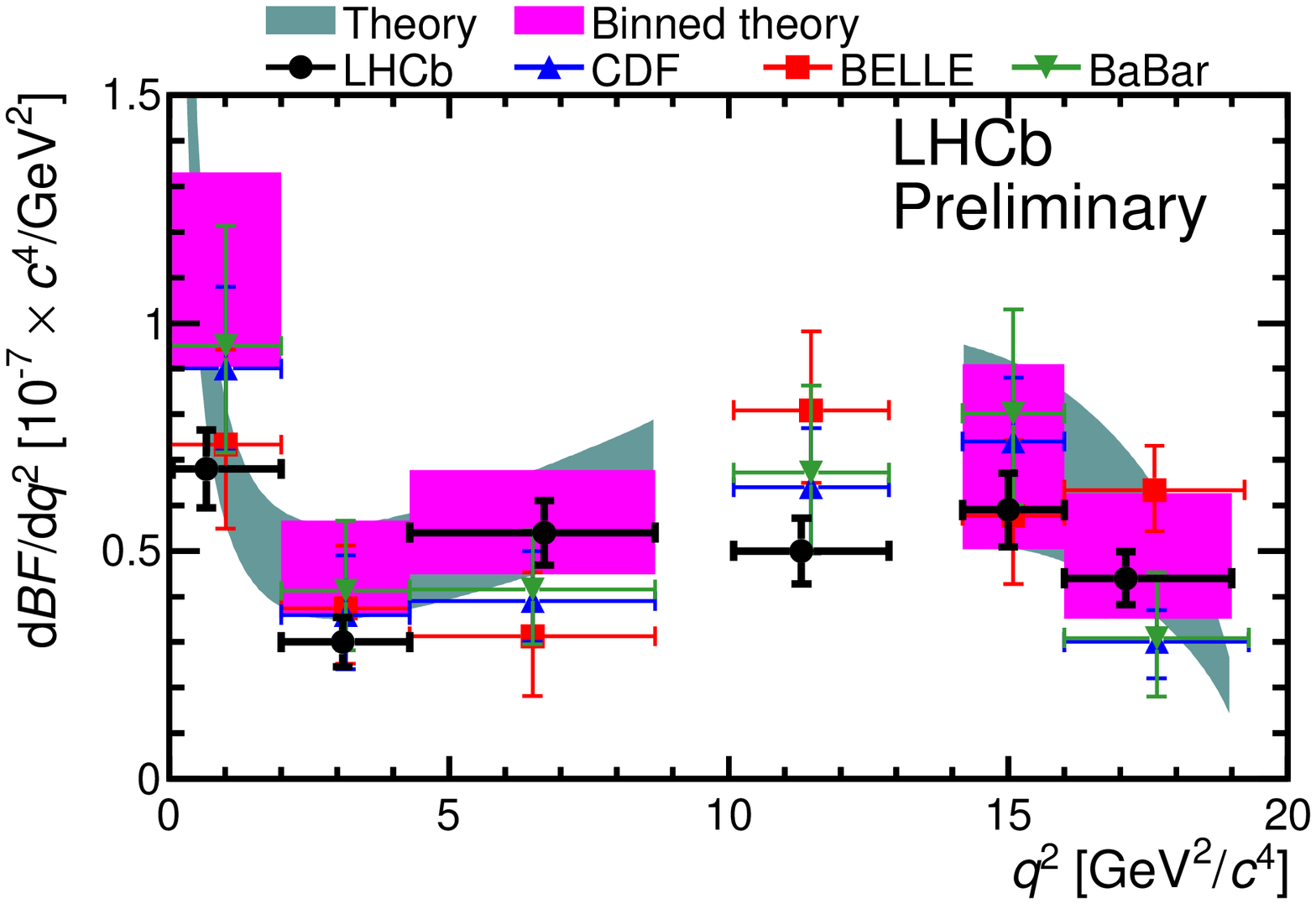} \\
\includegraphics[width=4.2cm]{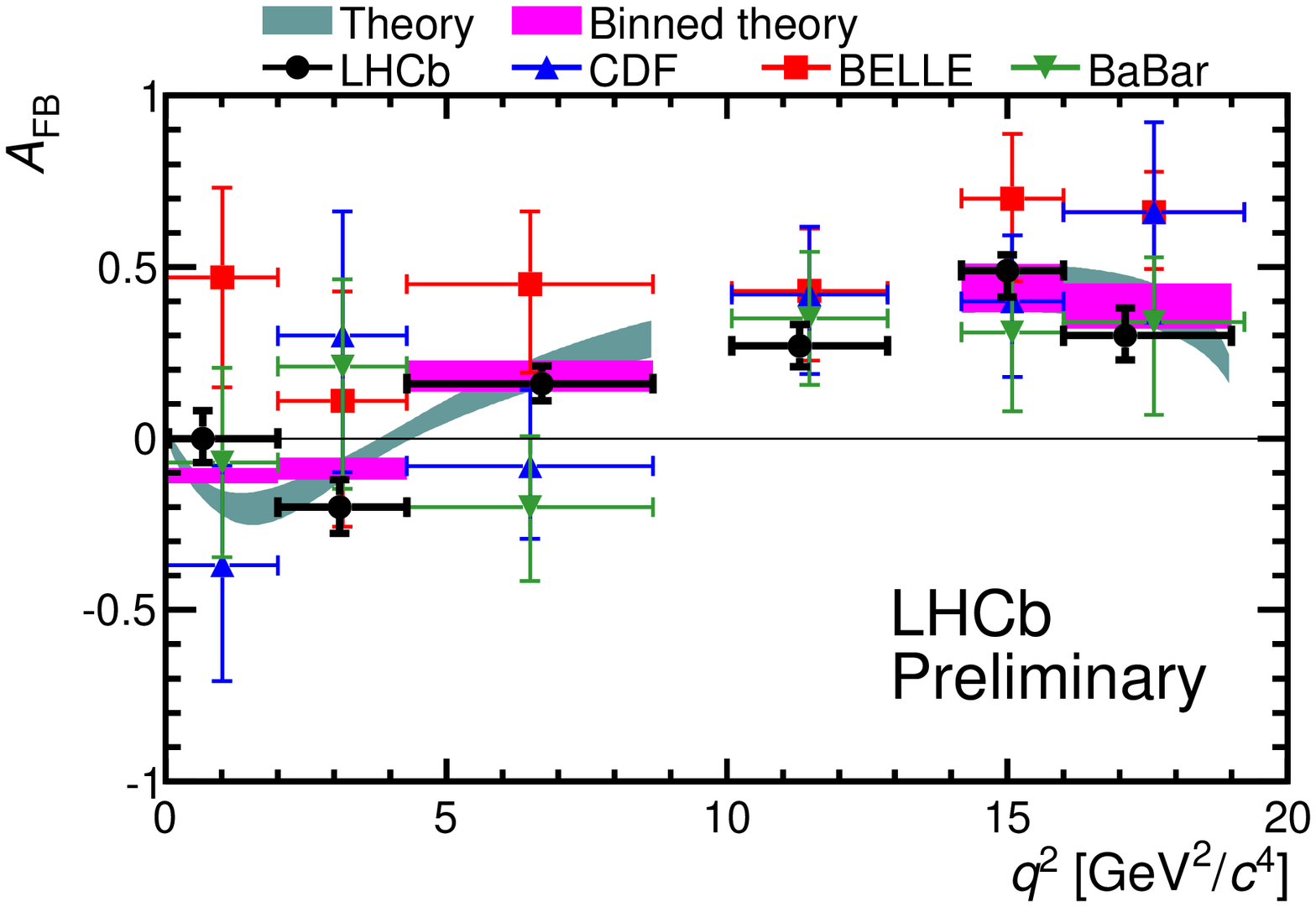} % Here is how to import EPS art
\includegraphics[width=4.2cm]{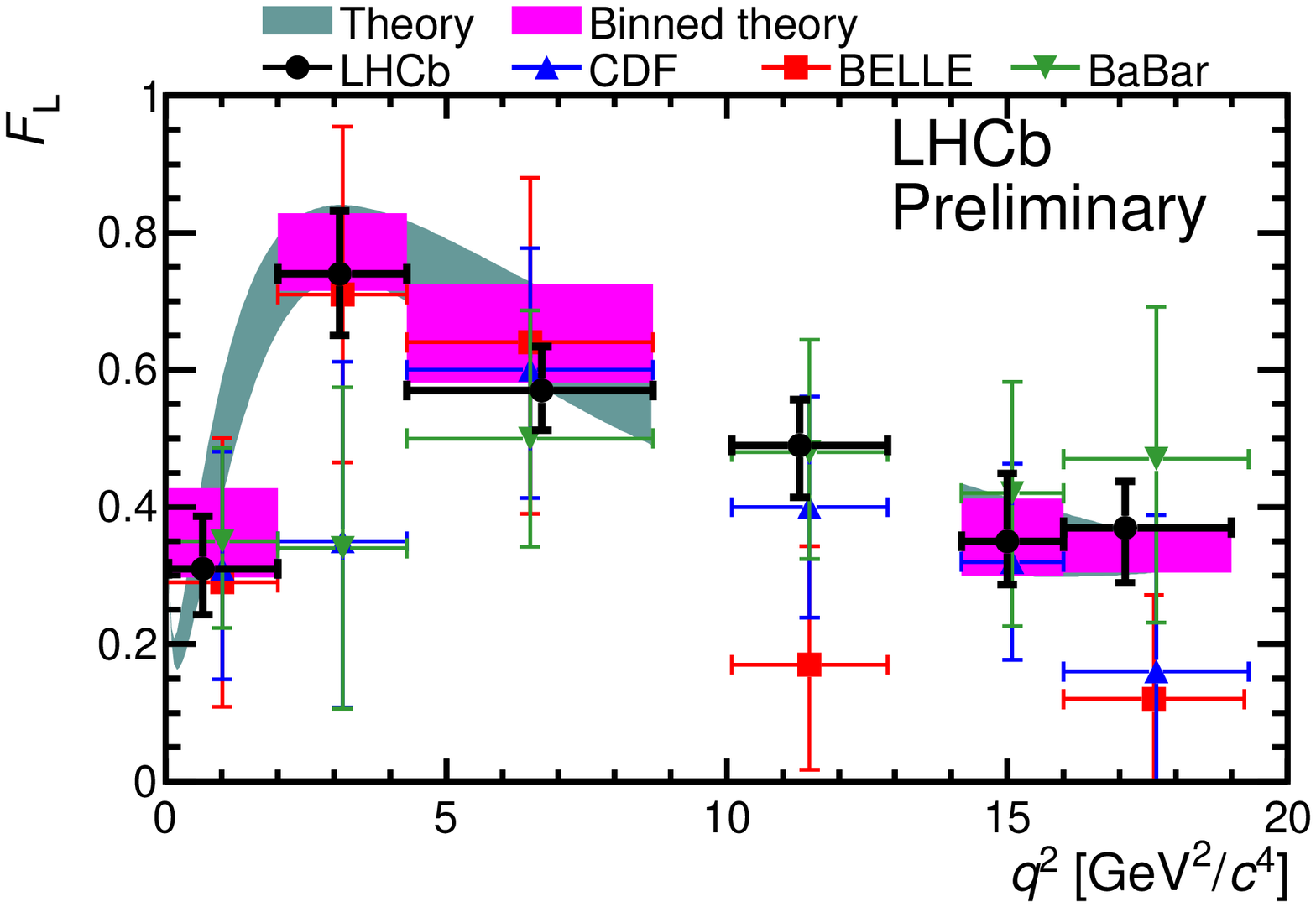}\\
\includegraphics[width=4.2cm]{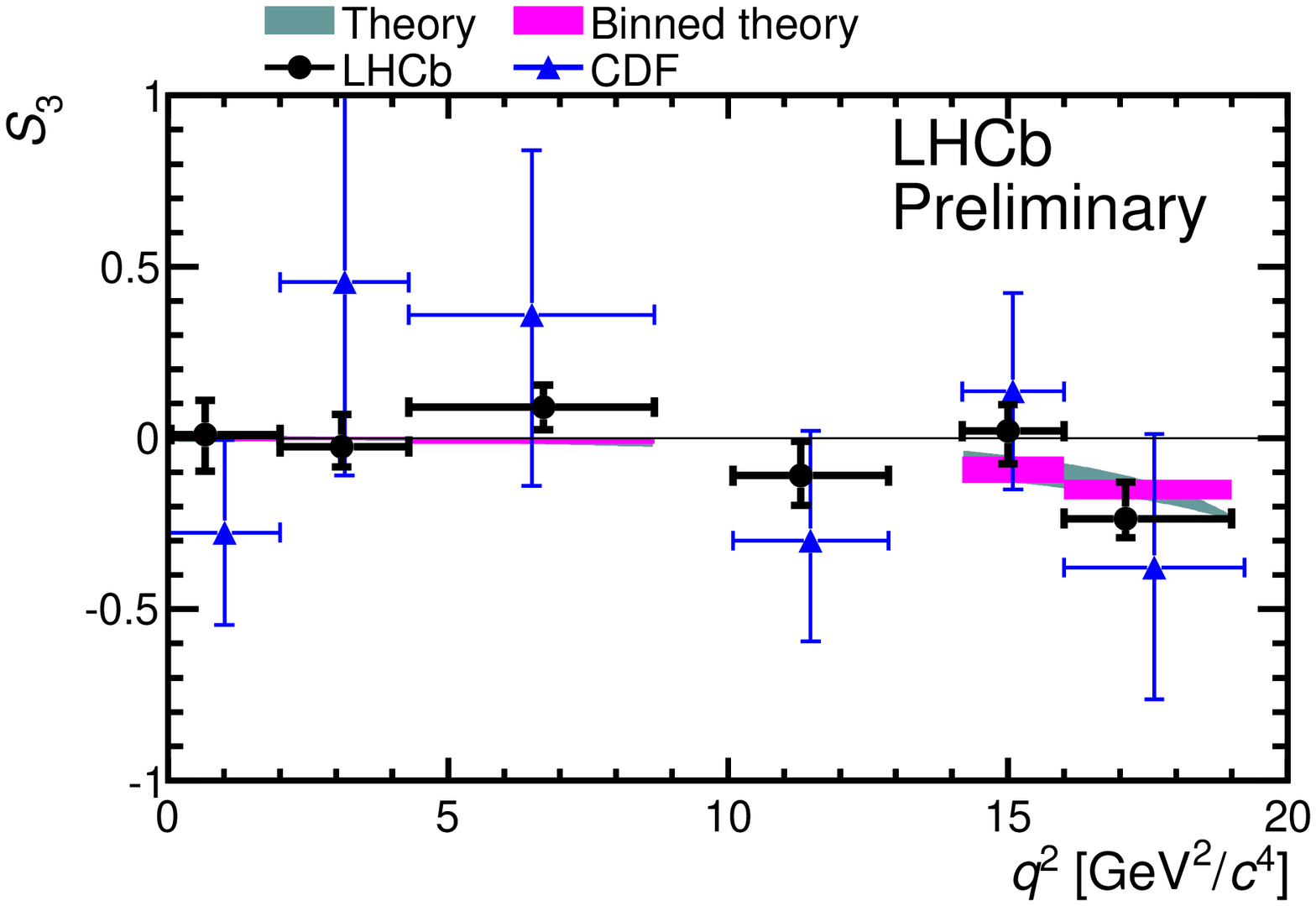}
\includegraphics[width=4.2cm]{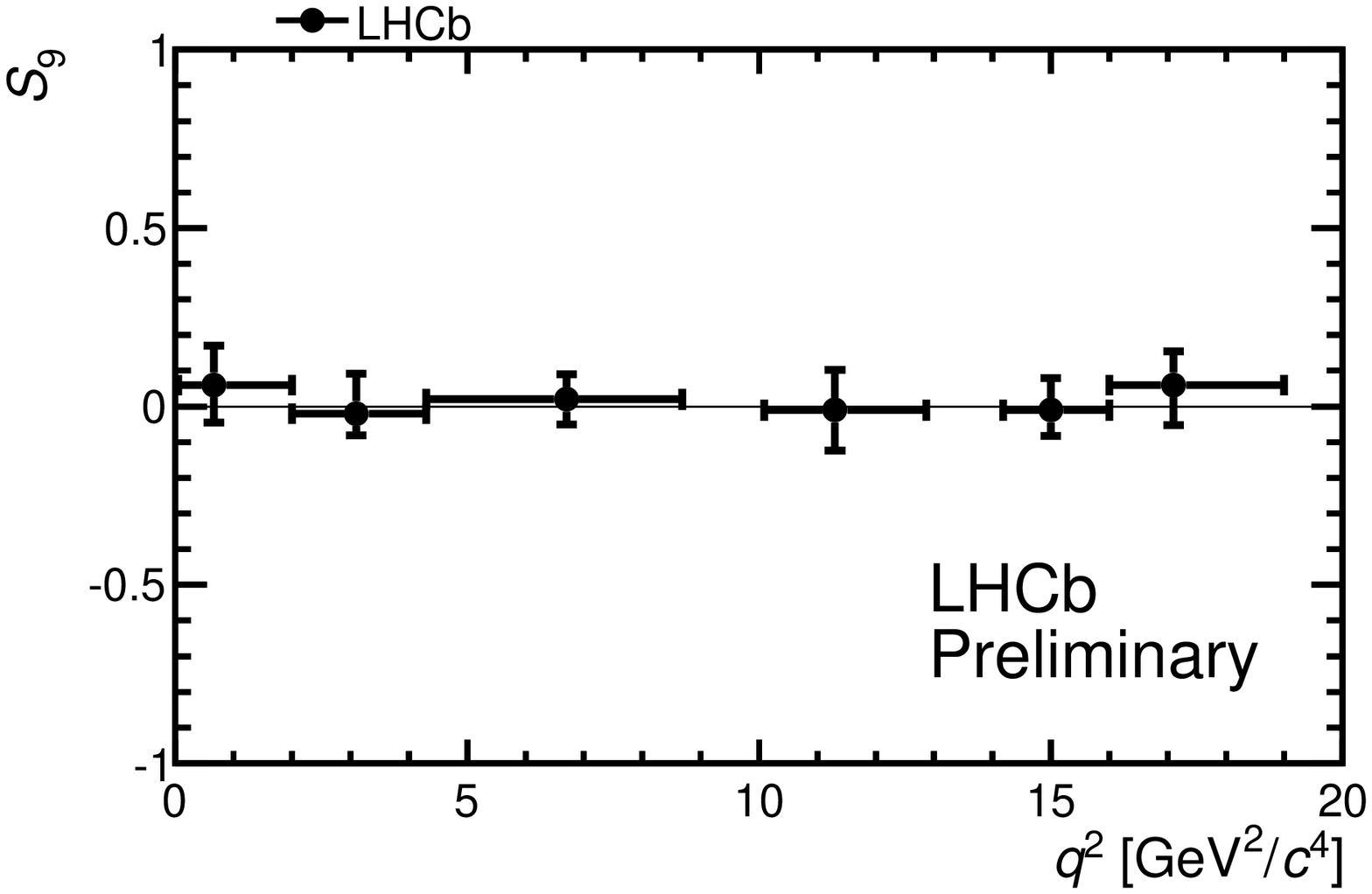}
\caption{\label{fig:BToKstarmumu} Results of the measurements of the $BR$, $A_{FB}$, $F_L$, $S_3$, $S_9$ as function of $q^2$ (black dots) compared to the theoretical predictions (colored curves) for the $B^0 \to K^{*0} \mu^+\mu^-$ decays.}
\end{figure}

%B->K*mumu angular
The helicity structure  of the $B^0 \to K^{*0} \mu^+\mu^-$ decay can be tested via the study of  the distributions of the helicity angle of the muon $\theta_L$, the helicity angle of the kaon $\theta_K$ and the angle $\phi$ between the decay plane of the $K^*$ and that of the di-muon system. These three angular variables, together with $q^2$, describe completely the kinematic of the decay. The differential decay rate as a function of the angular variables provides the sensitivity to some physics parameters, resulting from the combinations of  the amplitudes  in the transverse bases. They are the forward-backward di-muon asymmetry $A_{FB}$ , the fraction $F_L$ of longitudinal polarization of the $K^{*0}$,  $S_3$ related  to the polarization of the virtual photon giving the two leptons in one of the allowed diagrams (left handed in the SM), as well as $S_9$ related to the  imaginary part of the transverse amplitude. The experimental challenges for this analysis are the control of backgrounds  that could potentially pollute the angular distributions and the understanding of the biases induced by the geometrical acceptance of the detector. 
LHCb provides the most precise measurement  to date of all these observables and measures some of them for the first time, like $S_9$.  In addition, the branching fraction is measured as a function of $q^2$.  The results~\cite{kstarmumuang} are shown on figure~\ref {fig:BToKstarmumu}.
At the moment they are in agreement with the SM predictions. 

\begin{figure}
\includegraphics[width=8.4cm]{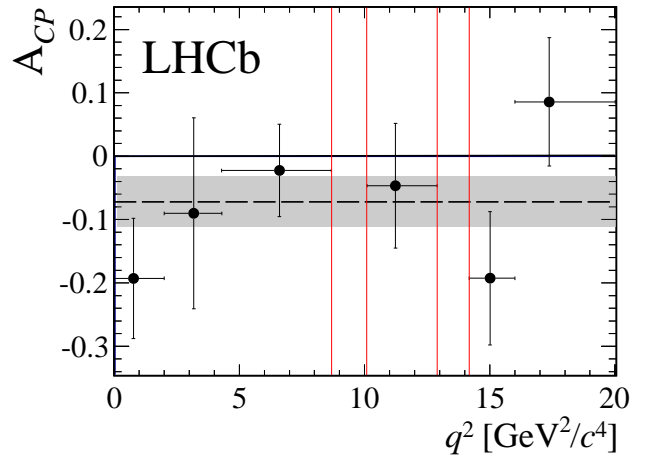} 
\caption{\label{fig:BToKstarmumuACP} $A_{CP}$ as function of $q^2$ for $B^0 \to K^{*0} \mu^+\mu^-$ decays.}
\end{figure}

%B->K*mumu Acp 
LHCb also measures the CP asymmetry in the  $B^0 \to K^{*} \mu^+\mu^-$ decays,   defined as: 
\begin{equation}\label{eq:Acp}
A_{CP}= \frac{\Gamma(\bar{B^0}\to \bar{K}^{*0}\mu^+ \mu^-) - \Gamma(B^0\to K^{*0}\mu^+ \mu^-)} {\Gamma(\bar{B^0}\to \bar{K}^{*0}\mu^+ \mu^-) + \Gamma(B^0\to K^{*0}\mu^+ \mu^-)}
\end{equation}
predicted to be of the order of $10^{-3}$ in the SM. Being defined as a ratio of amplitudes, uncertainties related to the presence of form factors cancel. 
The simultaneous fit measures the raw asymmetry $A_{RAW}=A_{CP}+kA_P+A_D$. To extract the $CP$ asymmetry, one can subtract that of the $B^0\to J/\psi K^*$ decay:  $A_{CP}=A_{RAW}(B^0\to K^{*0}\mu^+ \mu^-) - A_{RAW}(B^0\to J/\psi K^*) $. Indeed the $B$ production asymmetry $A_P$ is the same in the two decays, as well as the part of the detection asymmetry $A_D$ due to the different interaction with material of  particle with opposite charge. The part of $A_D$ due to the left-right detector asymmetry is taken into account averaging the measurements with magnet up and down.  Differences in the two decay kinematics are accounted for in the systematics by reweighing techniques. The 
different momentum distributions of $\mu^+$ and $\mu^-$ due to the $A_{FB}$ are studied with a $J/\psi$ control sample. The result is shown on figure~\ref{fig:BToKstarmumuACP}~\cite{KstarmumuACP} .

\begin{figure}
\includegraphics[width=4.2cm]{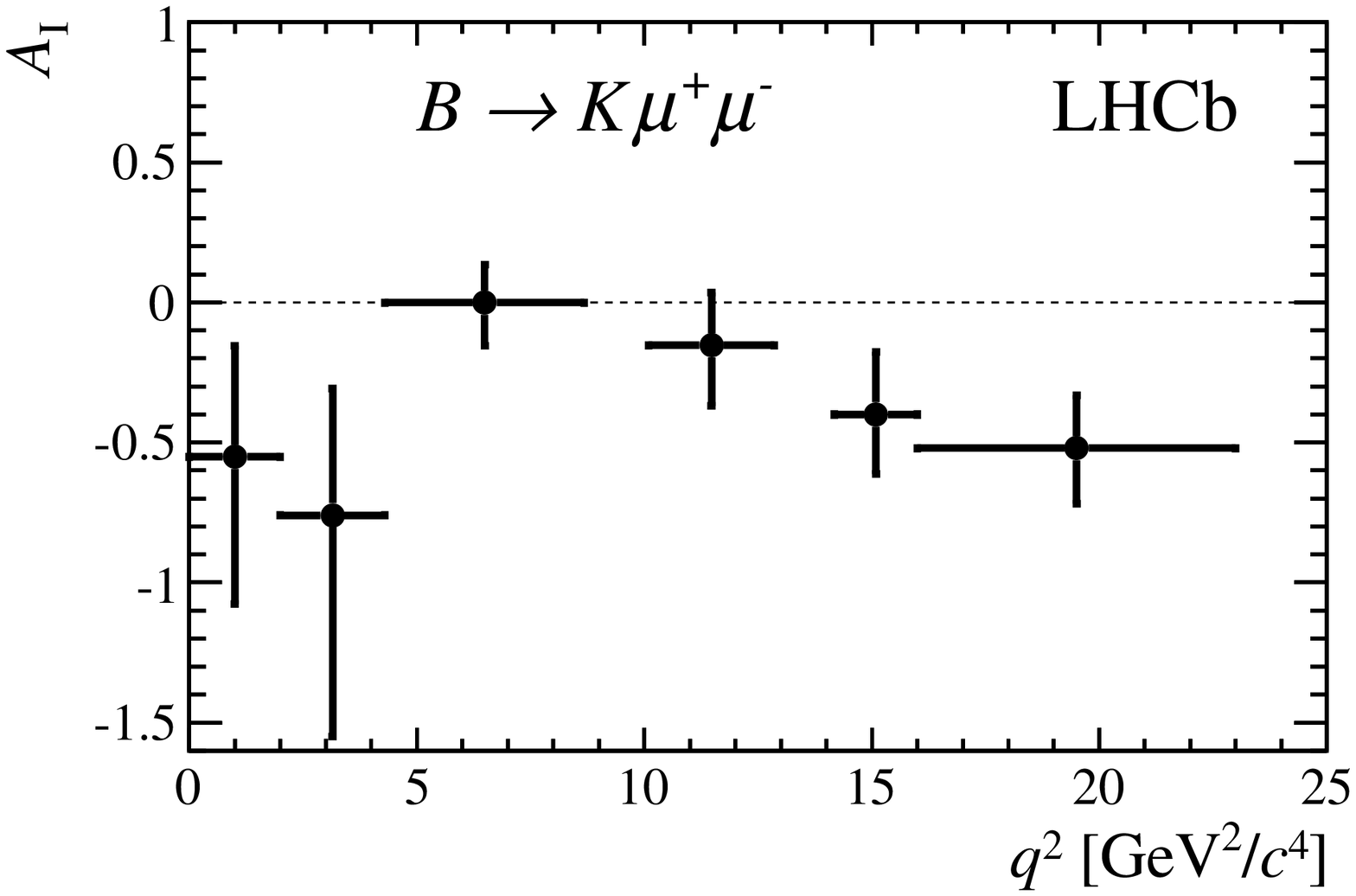} 
\includegraphics[width=4.2cm]{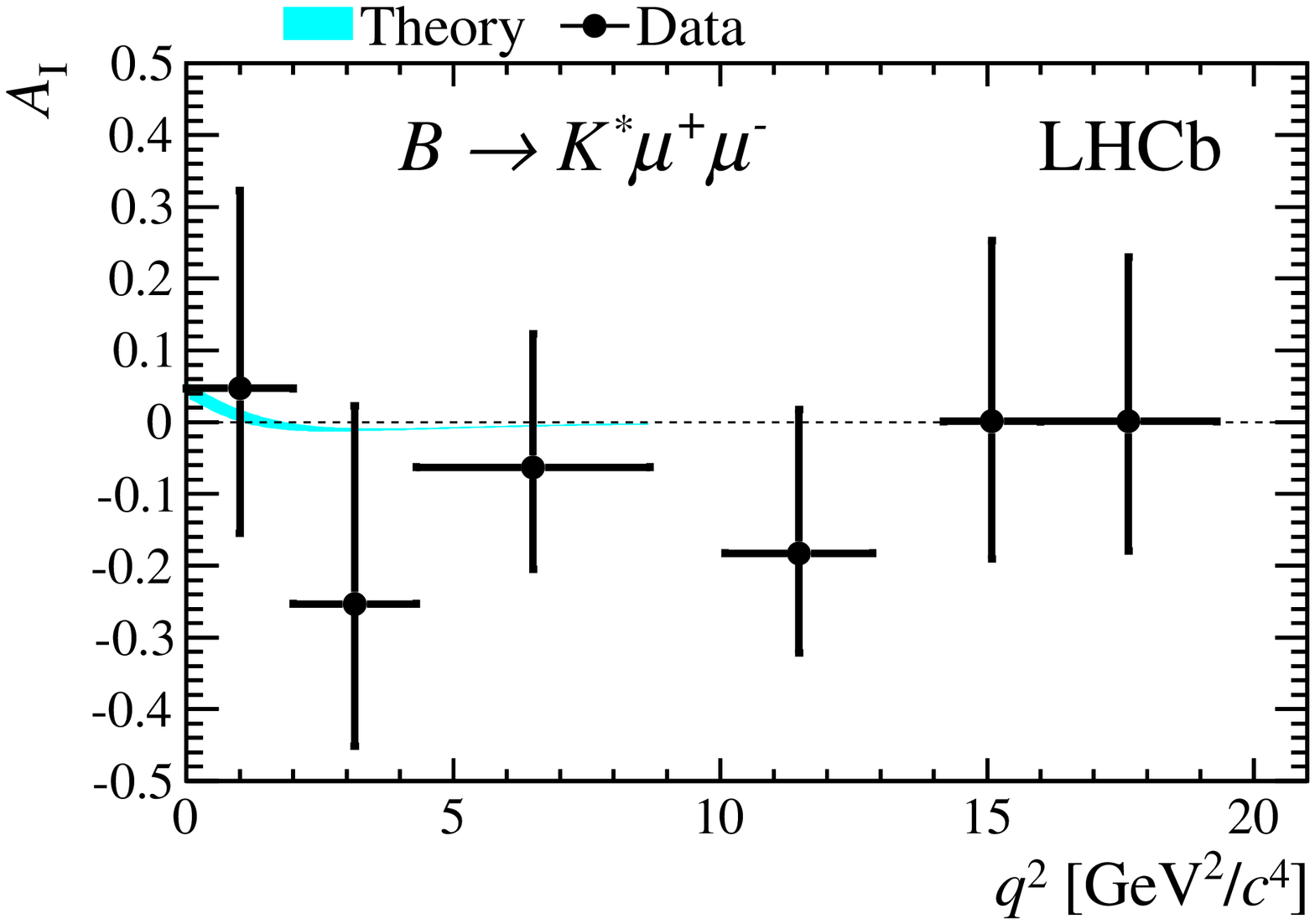} 
\caption{\label{fig:BToKstarmumuISO} The isospin asymmetry for the $B \to K \mu^+\mu^-$ (left) and the $B \to K^* \mu^+\mu^-$ (right) decays.}
\end{figure}

%B->K(*)mumu isospin
The $B \to K^{(*)} \mu^+\mu^-$ decays are finally used to perform an isospin analysis. The quantity which is measured is the isospin asymmetry defined as follows:
\begin{equation}\label{eq:Aiso}
A_I = \frac{\Gamma(B^0\to K^{(*)0}\mu^+ \mu^-) - \Gamma(B^+\to K^{(*)+}\mu^+ \mu^-)} {\Gamma(B^0\to K^{(*)0}\mu^+ \mu^-) + \Gamma(B^+\to K^{(*)+}\mu^+ \mu^-)}
\end{equation}
In the SM this is predicted to be extremely small, mainly due to the initial state radiation which is slightly different for the two different spectator quarks of the $B^+$ and $B^0$. A simultaneous fit to all the channels gives the results~\cite{Kstarmumuiso} shown in figure~\ref{fig:BToKstarmumuISO}. 
The $B \to K \mu^+\mu^-$ decays show a negative isospin asymmetry integrated over $q^2$, with a deviation of 4.4 $\sigma$ from zero. This effect is mainly due to a deficit in the $B^0 \to K^0_s \mu^+\mu^-$ decay, observed with 5.7 $\sigma$ significance. 
The $B \to K^* \mu^+\mu^-$ decays show a negligible isospin asymmetry as predicted in the SM. 
The branching fractions are measured to be $BR(B^0 \to K^0 \mu^+\mu^-) = (0.31^{+0.07}_{-0.06})\times10^{-6}$ and $BR(B \to K^* \mu^+\mu^-) = (1.16\pm 0.19)\times10^{-6}$. 

%B->K*ee 
The $B^0 \to K^{*0} e^+e^-$ decay is also a $b\to sll$ transition, complementary to the channel with muons.  Thanks to the lower mass of the electrons respect to the muons, lower $q^2$ are explored analyzing this decay, giving  a higher sensitivity to the photon polarization.  In addition, being the electron mass negligible, the formalism is simpler. On the other hand this channel as a worst resolution due to the electrons bremssthralung effect. LHCb has the first observation with 4.6$\sigma$ significance of this decay.    
The analysis is performed in the range $ 0.03 < q^2 < 1$ $GeV^2/c^4$ to avoid high contamination of $B \to K^* \gamma$ and the degradation of the $e^+e^-$ plane measurement  due to multiple scattering. The $B \to K^* J/\psi(e^+e^-)$ channel is used as control sample. Specific cuts are applied to reject backgrounds from $B^0 \to D^-e^+\nu$ decays with $D^- \to K^*e^- \bar{\nu}$ and from $B^0 \to K^* \gamma$ with the photon converting to an electron-positron pair.
The resulting branching fraction is: $BR(B^0 \to K^{*0} e^+e^-)=(3.1^{+0.9 +0.2}_{-0.8 -0.3}\pm0.2) \times10^{-7}$~\cite{Kstaree} . An angular analysis is foreseen for the future.

%radiative decays B->K*gamma and B->Phi gamma
The radiative decays  $B \to K^{*} \gamma$  and $B_s \to \phi \gamma$ are also interesting probes to new physics. The SM predicts the two decays to have the same branching fractions of $(4.3 \pm1.4) \times 10^{-5}$ and a CP asymmetry $A_{CP}(B^0 \to K^* \gamma)=(-0.61 \pm 0.43) \%$.  The LHCb analysis~\cite{Aaij:2012ita} founds: $BR(B \to K^{*} \gamma)/BR(B_s \to \phi \gamma)=1.23 \pm 0.06_{stat.} \pm 0.04_{syst.} \pm 0.10_{f_s/f_d}$ and $A_{CP}(B^0 \to K^{*} \gamma) = (0.8 \pm 1.7_{stat.} \pm 0.9_{syst.})\%$, in agreement with the expectations.

\section{Conclusions}
Rare decays are sensitive probe to new physics effects. 
The results of recent  searches performed by the LHCb experiment using 1.0 $fb^{-1}$
of $pp$ collisions collected by the experiment in 2011 at a centre-of-mass energy of $\sqrt{s}$=7 $TeV$ do not show evident signs of new physics effects.  Nevertheless the experiment is strongly pushing forward previous limits, giving the most precise measurements to date, and observes very rare decays never observed before. For some decays it has collected enough data for a complete analysis of their properties, thus providing the informations to further constrain the possible new physics contributions. 

%\subsection{References}

%\begin{acknowledgments}
%We wish to acknowledge the support of the author community in using
%REV\TeX{}, offering suggestions and encouragement, testing new versions,
%\end{acknowledgments}

%\nocite{*}


\input{bibliography}

\end{document}

%% file: bibliography.tex
%\newpage
%\thispagestyle{empty}
%\
%################ LIVROS #################
%\scriptsize

%\newpage
%\thispagestyle{empty}
%\
%\\